# Learnt Microwave Image Reconstruction with A Conformal Antenna Array

Wenyi Shao and Beibei Zhou

***Abstract*—A deep learning model is proposed for reconstructing 2D dielectric breast images from time-domain signals. Unlike existing learning models that employ a fixed antenna array, where input data consists solely of measurements, the proposed system integrates antenna positioning into the processing pipeline. This allows for a conformal antenna array that adapts to different breast sizes for optimal data collection across various patients, which eliminates undesired signal attenuation in coupling liquid when implemented for the fixed array. By leveraging antenna positions, the breast surface can be pre-estimated, enabling the neural network to focus on image reconstruction within the region of interest. Numerical results demonstrate that the proposed model may reconstruct breast images with good quality.**

## I. Introduction

Microwave breast imaging (MBI) is a promising technology for routine breast screening, offering the advantages of being non-ionizing and cost-effective [1–9]. Traditional MBI algorithms can be broadly categorized into two groups. The first focuses on reconstructing dielectric properties within the breast. Representative techniques include nonlinear iterative methods such as contrast source inversion (CSI) [10] and distorted Born iterative method (DBIM) [11]. The second group consists of radar-based methods, which only aim to localize tumors [12-26] and are generally only effective for imaging less-dense breasts.

Deep learning (DL) has been recently applied across various medical imaging modalities [27–32], enhancing image quality by improving contrast and spatial resolution. It has also been incorporated into microwave medical imaging to classify tumors, either based on reconstructed images via traditional methods [33-35] or directly on microwave measurement data [36]. In addition, DL has also been specifically developed for microwave breast image reconstruction displaying permittivity or conductivity profiles. In [37] and [38], training data were generated by first creating elliptical breast phantoms with axes ranging from 6.5 to 12 cm. These phantoms were then populated with randomly assigned tumor, fibro-glandular, transitional, and adipose tissues and enclosed by a skin layer 1.5–2.5 mm thick. Microwave data were simulated using 30 antennas arranged in a circular configuration with a fixed 12 cm radius, operating at a single frequency.

Most published DL studies employ frequency-domain data for image generation [37-40], utilizing either single or multiple frequencies. However, time-domain signals inherently contain broadband information, offering the potential for high-resolution image reconstruction. Despite this advantage, direct mapping between time-domain data and the image domain remains largely unexplored. To date, we have identified only two machine learning (ML) approaches that utilize time-domain signals. In [36], time-domain signals were used to train an ML model to classify tumors as benign or malignant, but the tumor location needed to be pre-known. In [41], a U-Net model was trained on time-domain signals using transfer learning to reconstruct brain profiles. While U-Net has demonstrated strong performance in image segmentation and denoising, its structure is less suited for direct image reconstruction from raw signals. The primary limitation arises from U-Net's reliance on skip connections, which preserve spatial features by linking early and later layers at corresponding scales. This architecture is effective when a spatial correspondence exists between the input and output, as in segmentation tasks, both belong to the image domain and typically have the same dimensions. However, in signal-to-image conversion, the input (signal domain) and output (image domain) belong to fundamentally different representations, making the direct use of U-Net suboptimal without premodifications. As an example, to generate 256×256-pixel images, literature [41] utilized 256 signal channels and downsampled the original 5,000 time-step signals to 256 time steps. This downsampling is only to facilitate network training, but may result in information loss, potentially sacrificing spatial resolution.

This paper proposes a preliminary study for using learning-based approach to map time-domain microwave data to the image domain for breast imaging. To generate training data, a pair of antennas—one transmitter and one receiver—are positioned at 24 evenly spaced azimuthal locations around the breast phantoms, operating in a full multi-static mode. This setup provides a cost-effective alternative to a full antenna array while ensuring adequate data acquisition coverage. While liquid coupling media are widely used in MBI due to several advantages [42–44], they also present significant drawbacks: The typical setup involves a tank filled with liquid to immerse both the breast and antennas, introducing procedural complexities such as pre-filling, bubble elimination, temperature regulation, and post-examination cleaning. Secondly, liquid disposal raises environmental concerns, and reusing the medium across multiple patients may pose hygiene risks. More importantly, immersion in liquid can cause discomfort, strongly affecting the patient's examination experience. An alternative approach is direct antenna contact, where customized antennas are pressed against the breast. However, pressure from rigid antennas may also cause discomfort, and even with firm contact, achieving perfect conformity is still challenging. Local air gaps may introduce impedance mismatches and degrade signal transmission. Firm compression further alters the breast shape. Since each antenna position requires a new measurement, the breast may deform differently at each step. These shape variations introduce inconsistencies in the imaging process, complicating reconstruction and reducing overall accuracy.

To address these challenges, recent advancements propose semi-solid materials as an alternative matching medium [45]. These ultra-soft, polyester-based materials can be reshaped to conform to the breast surface, minimizing breast deformation while contacting and also reducing impedance mismatch. Building on this concept, we introduce a conformal-antenna measurement system that integrates semi-solid matching material and incorporates this feature into neural network (NN) development. The key innovations of this work include:

(1) Adaptive Antenna Configuration: Unlike all other DL-based imaging systems that use fixed-radius circular arrays [33-41, 46-48], our system features an adaptive antenna configuration. Each antenna is embedded in a semi-solid matching medium and moves radially [49-52] to establish direct contact with the breast, accommodating various breast sizes. This design minimizes signal loss in coupling liquid associated with fixed-radius arrays, particularly for smaller breasts.

(2) Integration of Antenna Coordinates to NN: The adaptive antenna position is incorporated into the NN through an antenna-coordinate input channel. A transformer model (TM) [53] processes this spatial information, converting it to a feature vector fed to the image reconstruction network. To the best of our knowledge, this is the first learning-based approach integrating antenna locations for near-field imaging.

(3) Breast Surface Estimation**:** Since the antennas make direct contact with the breast, their positions can be utilized to estimate the breast surface profile. This profile is then passed to the NN, allowing it to focus solely on the region within the profile, thereby enhancing image reconstruction quality and accelerating NN convergence during training.

(4) Robust Training Data Generation: To train the NN, a large dataset of synthetic breast phantoms resembling real breasts was generated by a generative adversarial network (GAN) [54-60]. These phantoms provide diverse and realistic training samples, improving model robustness.

The system design and NN architecture are detailed in Section 2, followed by imaging results in Section 3. Comparative analyses and discussions are presented in Section 4, with final conclusions provided at the end of the paper.

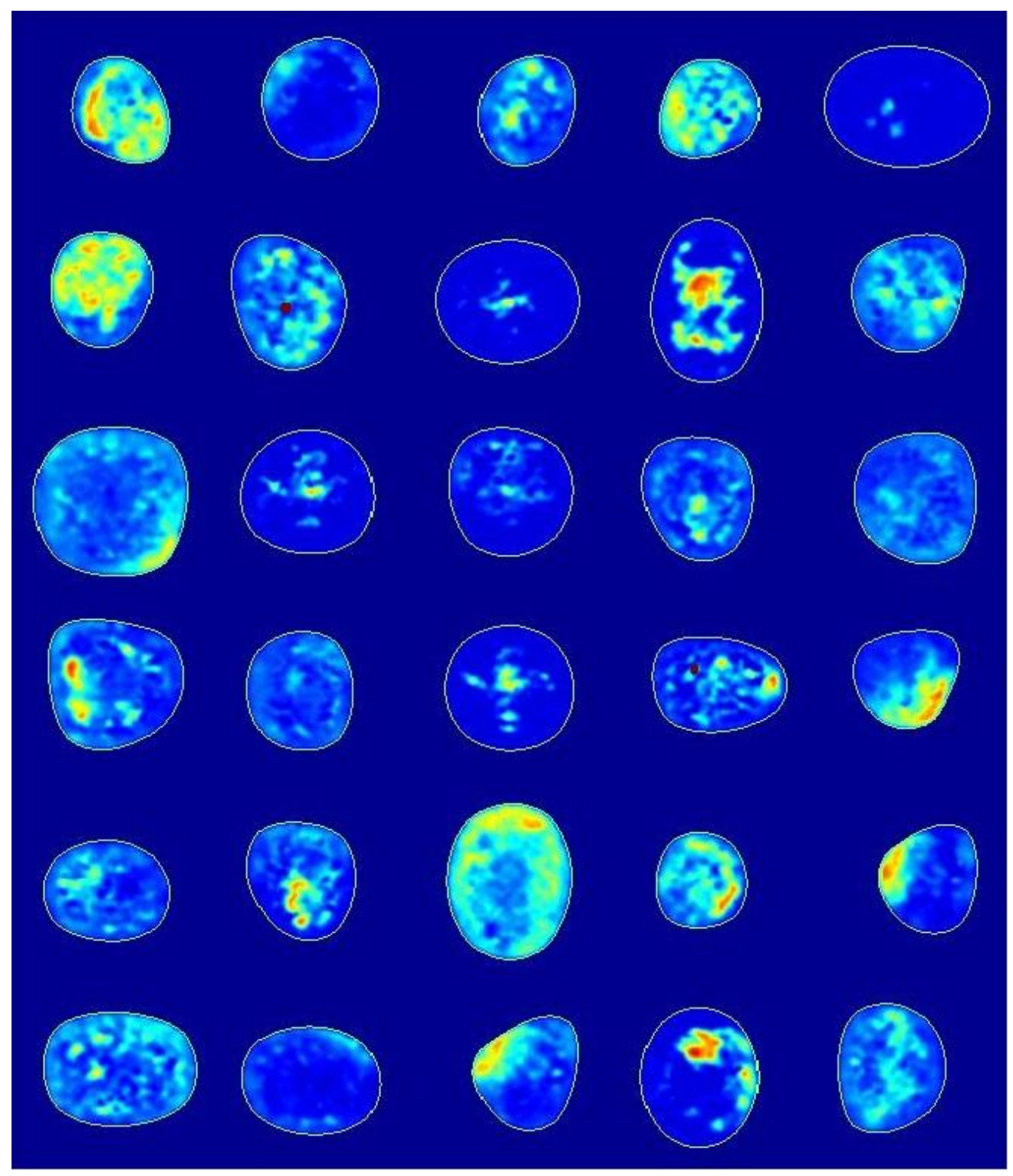

Fig. 1. Generated 2D digital breast phantoms by GAN.

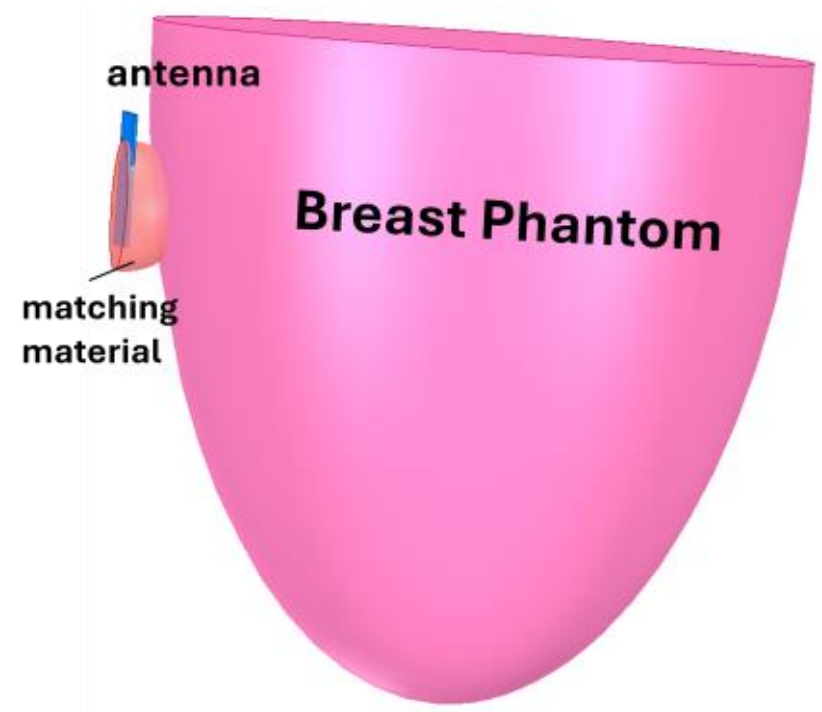

Fig. 2. Antenna wrapped in semi-solid matching material that conforms to the breast surface. Figure is intended for conceptual visualization only. The current study focuses exclusively on 2D imaging.

## II. Method

### A. *Synthetic Breast Phantoms*

There have been reports of researchers fabricating a few hundred simplified phantoms and using a vector network analyzer (VNA) to collect measurement data for ML application [61]. However, such limited datasets increase the risk of overfitting, as ML models typically require large amounts of diverse data to generalize effectively. State-of-the-art ML architectures— such as InceptionV4, GoogleNet, and ChatGPT— are trained on massive datasets, often comprising tens of millions of samples, to achieve high performance in tasks such as image recognition or generation. In the context of medical imaging, this implies the need for a large, heterogeneous patient dataset to serve as ground truth. Yet, acquiring such a dataset in practice remains a significant challenge.

To address this limitation, we generated synthetic breast models using a previously trained GAN [54, 56, 59]. This generative model can produce 128×128 2D digital breast phantoms (permittivity and conductivity maps in pair at 1 mm resolution) that closely resemble real patient data. Until a sufficiently large clinical dataset becomes available, leveraging synthetic phantoms along with their corresponding measurement data — generated via efficient simulation methods — provides a practical solution for NN training. In this study, we generated 20,000 breast phantoms, which were used both to simulate microwave measurement data and to serve as ground truth for training the NN in image reconstruction. Examples of the generated permittivity maps are shown in Fig. 1. To simulate early-stage tumor occurrence, small elliptical or spherical tumors (≤1 cm in diameter) were incorporated in approximately 12% of the phantoms, consistent with the statistic that roughly 1 in 8 individuals may develop breast cancer during their lifetime. The dielectric properties of the tumors were set 10% higher than the maximum value found in fibroglandular tissues in each individual phantom.

Before being used for training, the dielectric values of all phantoms were normalized by the global maximum value across all 20,000 phantoms, i.e.,

$$I' = \frac{I}{\max\,(I_{20000})} \tag{1}$$

where $I$ is the permittivity or conductivity before normalization, $\max\,(I_{20000})$ denotes the highest permittivity or conductivity value across the entire dataset (typically that of a tumor), and $I'$ is the normalized value, constrained within the range (0, 1].

### *B. System Description*

Consider a UWB modified monopole antenna [62-67] is embedded within a semi-solid matching material that maintains direct contact with the breast, as illustrated in Fig. 2. Two such antennas—one serving as a transmitter and the other as a receiver—are positioned to move both azimuthally and radially [50] around the breast, at 24 azimuthally evenly spaced locations, forming a virtual antenna array as depicted in Fig. 3. In this configuration, antennas (represented by black dots) are spaced 15° apart in azimuth. Each antenna can move radially along the red dashed lines until it reaches a position 5 mm from the breast surface, ensuring optimal coupling when the matching material conforms perfectly to the breast.

Based on the actual antenna locations, we compute 24 corresponding inner points (shown as blue dots in Fig. 3) by shifting each antenna position 5 mm inward toward the origin. Assuming these blue points lie on the breast surface, we estimate the breast boundary by fitting an ellipse (blue curve) to these points using the least-squares (LS) method, which minimizes the total distance between the fitted ellipse and the blue points $(x_i, y_i)$:

$$\min_{ellipse} \sum_{i=1}^{24} \min_{1 \le j \le N} \left\| (x_i, y_i) - (x_j, y_j) \right\|^2 \tag{2}$$

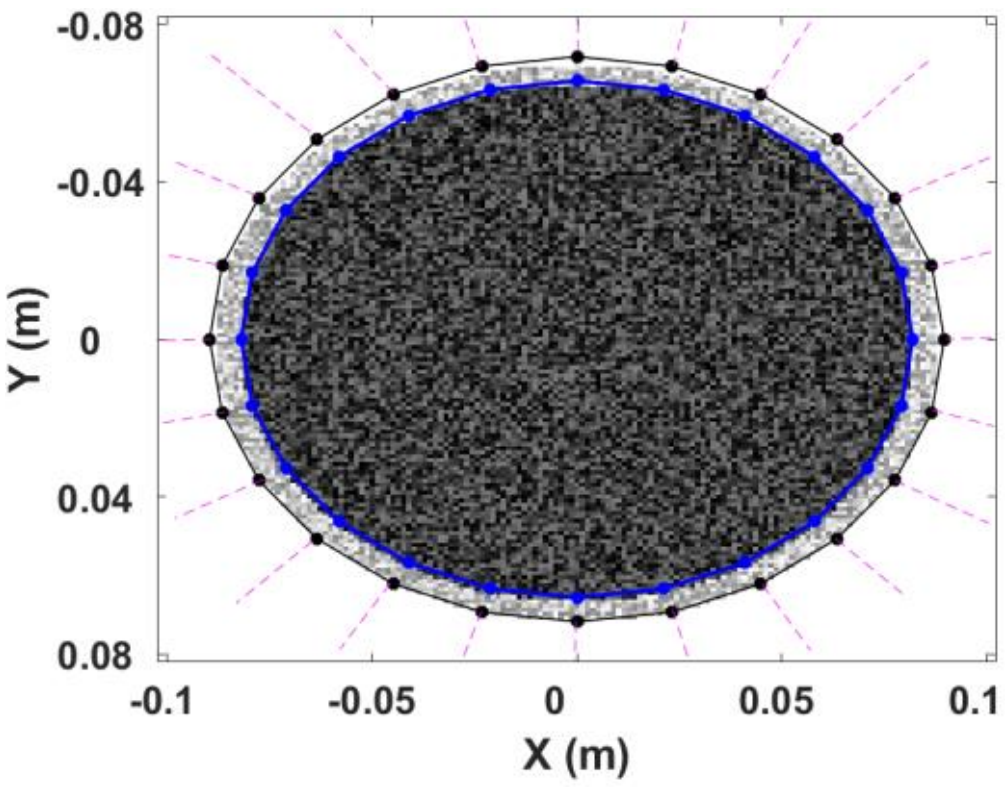

Fig. 3. The antenna positions and the probability map. Antenna can move radially until the matching material (matching material not shown in the figure) conforms perfectly to the breast surface.

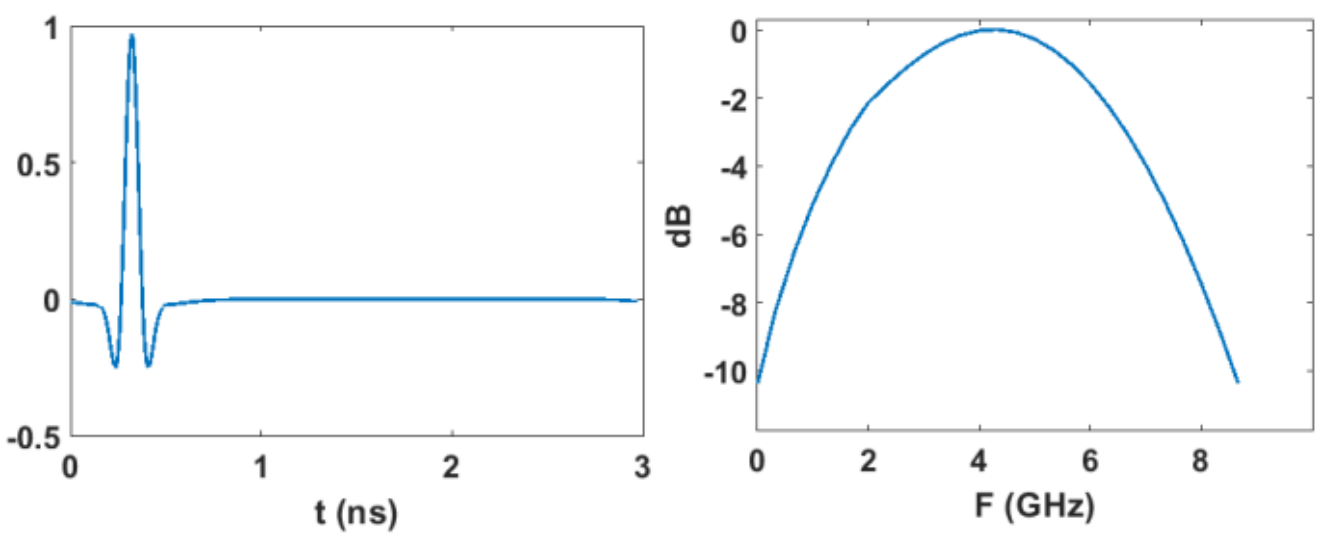

Fig. 4. The time-domain excitation (left) and its spectrum (right).

where $(x_j, y_j)$ are the pixel coordinates of $N$ points on the discretized ellipse.

Since image reconstruction will primarily focus within the estimated boundary, we assign high probability values (randomly ranging from 0.7 to 0.95, shown as dark pixels) to the corresponding region. However, considering that some breast tissue may extend beyond the estimated boundary, we also assign lower probability values (randomly between 0.2 and 0.7, shown as light gray pixels) to the area between the blue ellipse and the black polygon formed by connecting all antenna locations. Outside this polygonal boundary—i.e., beyond the antenna array—breast tissue is unlikely to be present, and thus the probability is set to zero (shown as white). This spatial probability map is represented on a 128×128 grid, where each pixel corresponds to 1 mm in physical space. By providing this map as an input to the NN, the model is guided to concentrate its reconstruction efforts within the region of interest while disregarding areas with zero probability.

### *C. Simulation*

With the generated breast phantoms, we employed finite-difference time-domain (FDTD) simulations to obtain time-domain electromagnetic responses, utilizing a parallel algorithm to accelerate computation [68-74]. The simulations took approximately 10 hours on an 8-core Intel i7 workstation. Although DL-based forward modeling methods have been explored [75], their accuracy compared to conventional approaches remains an active area of research. Therefore, we adopted the traditional FDTD method for this study to ensure

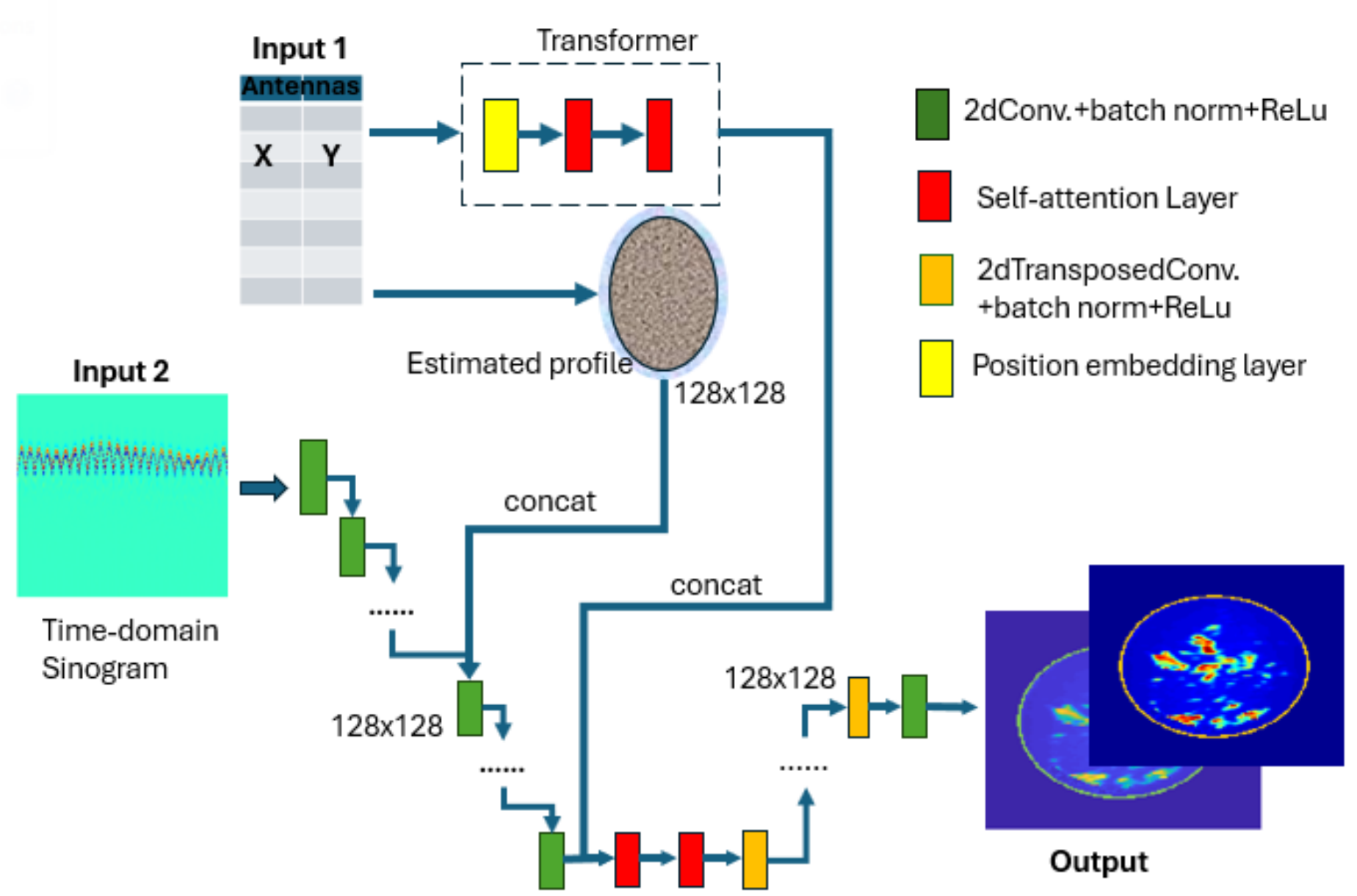

Fig. 5. Architecture of NN for reconstruction.

simulation fidelity.

The excitation signal was an ultra-short pulse with a −3 dB bandwidth spanning approximately 5 GHz (2–7 GHz), as illustrated in Fig. 4. During each simulation, data were sampled at 5 picosecond intervals, resulting in 585 time steps per recording. Given the 24 antenna positions operating in full multistatic mode, each simulation generated 552 signals (i.e., 24×23), forming a 585×552 sinogram. This sinogram serves as the input to the NN via the signal channel.

### D. NN Architecture and Training

The NN architecture is illustrated in Fig. 5, and designed based on the following concepts:

(1) **Signal Processing Pipeline:** The sinogram (585 × 552) is processed using a 2D convolutional NN (CNN) to extract features both in the time domain and across the antenna elements.

(2) **Antenna Location Processing:** A 24 × 2 coordinate matrix (representing the X and Y positions of the antennas) is fed into a transformer model (TM) to encode spatial information.

(3) **Probability Map Integration:** A probability map, derived from array geometry and the LS algorithm, is incorporated to guide the reconstruction process.

(4) **Feature Fusion:** Signal features, antenna position data, and the probability map are fused to integrate physics-informed priors into the image reconstruction process.

(5) **Decoder:** A decoder module uses up-sampling and convolution to map the fused features onto a 128 × 128 output, representing the breast's dielectric parameter distribution.

The signal processing pipeline progressively reduces the feature map size to 128 × 128 before concatenation with the probability map. Convolutions along the time axis capture features such as skin reflections and scattering from the dense tissues and tumors (if present) in the time domain. Convolutions along the antenna axis capture interactions between neighboring antennas, enhancing spatial feature representation. After integrating with the probability map at the 128 × 128 level, additional 2D convolutional layers are applied to extract higher-level features. These are progressively compressed into a 1024 × 1 feature vector at the bottleneck of the network, which is then fused with the antenna-position encoding. The encoding process (downscaling) in total includes 11 2D-convolution layers, batch normalization, and ReLu function.

The TM, employed to process antenna positions, is able to effectively learn spatial relationships among the antennas even though they are irregularly distributed. The TM includes a position embedding layer, which encodes spatial coordinates into vector features [76], followed by two self-attention layers [53]. The self-attention mechanism computes attention scores to model dependencies between antennas and extracts meaningful spatial features, ensuring the network captures both absolute and relative positions of the antennas. The output of the TM is encoded antenna-position information, a 1 × 1024 vector, to be incorporated into the sinogram-signal-processing network.

The antenna position information and signal features are then fused to generate images through a series of transposed convolution and 2D convolution operations. The upscaling includes two transposed convolution layers followed by batch normalization and ReLu function, and a 2D convolution layer. The output consists of two images: one representing the relative permittivity and the other depicting the conductivity distribution, both for the 4.5 GHz frequency — the central frequency of the excitation signal.

Training was performed on an Nvidia GeForce RTX 3090 GPU for approximately 3.5 hours, using a standard backpropagation (BP) algorithm. Training employed a conventional mean squared error (MSE) loss function, with a learning rate of 0.0005. The network was trained for 100 epochs (9,000 iterations with a minibatch size of 200), as the loss plateaued beyond that point.

## III. Results

The dataset used for validation and testing was obtained from a separate source — the UWCEM Phantom Repository [77], which contains nine MRI-derived numerical dielectric breast phantoms based on real patient data. From this dataset, 424 horizontal slices were extracted, downsampled to a 1 mm grid resolution, and cropped to 128 × 128 pixels. To ensure consistency with the training configuration, the same virtual array setup was used to simulate the microwave signal responses. Using a completely different data source for testing offers a more objective evaluation of model generalization. The NN reconstructed all 424 breast slices in under one second.

Fig. 6 shows examples of the NN-generated permittivity and conductivity images for three representative cases: fibroglandular (1st row), dense (2nd row), and very dense (3rd row) breast phantoms. Pixel values were denormalized from the (0, 1] range back to their respective physical scales. For comparison, the ground truth and results from a traditional time-domain distorted Born iterative method (DBIM) [78] are also presented for the permittivity profiles. To ensure a fair comparison, the same probability map used in the NN was also supplied to the DBIM algorithm. In this setup, the DBIM reconstruction was constrained to the region with a nonzero probability, while the remainder of the image was set to air ($\varepsilon_r$ = 1.000006, conductivity = 0). Although this reduced the number of unknowns and slightly accelerated the DBIM process, the method remained computationally intensive—requiring approximately 1 hour to complete 24 iterations per case due to the need of FDTD-based forward modeling in each iteration.

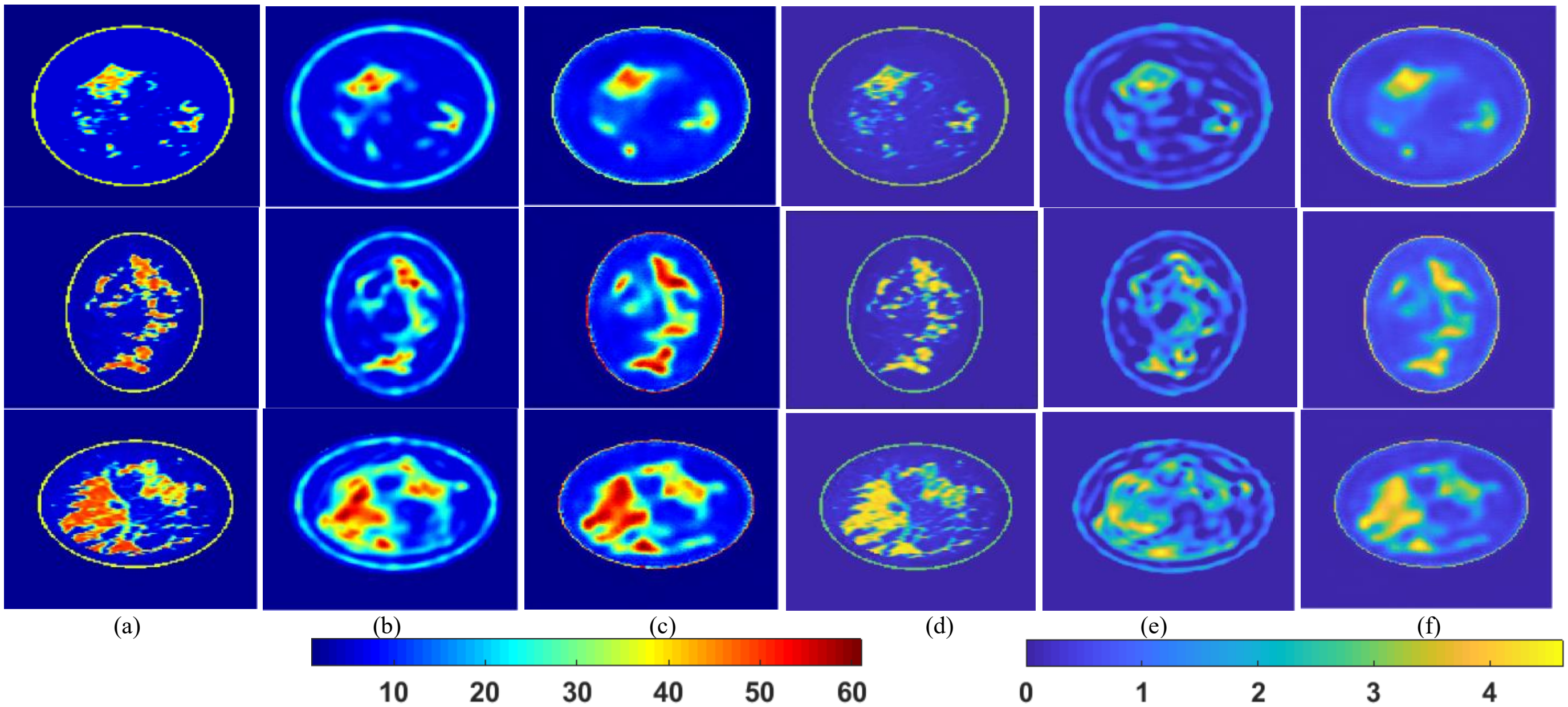

Fig. 6. Ground truth and reconstruction breast images for threes cases from the upper row, a least dense phantom, to the bottom row, a most dense phantom. (a) Ground truth relative permittivity of phantoms; (b) Reconstructed relative permittivity by DBIM; (c) Reconstructed relative permittivity by NN; (d) Ground truth conductivity (S/m) of phantoms; (e) Reconstructed conductivity by DBIM; (f) Reconstructed conductivity by NN.

At first glance, both reconstruction methods accurately reflect the general fibroglandular pattern of the breast. DBIM produces smoother spatial variations across neighboring pixels, while the NN reconstruction reveals finer structures, such as the skin layer, with greater clarity. This implies a super-resolution capability, as has been noted in previous literature. However, careful inspection reveals that the skin boundary in the NN reconstruction does not exactly match that in the ground truth. One may question that the NN might just add a skin layer around the breast tissues, learnt from the training-phantom dataset. Notably, the dielectric values within the reconstructed skin layer vary across different regions—some are higher, others lower—compared to the uniform skin values in the ground truth. If the NN were merely memorizing the presence of a skin layer, it would more likely reproduce it with a consistent dielectric value. The observed inconsistency indicates the NN is not copying, but rather inferring the dielectric values in and near the skin layer.

TABLE I
COMPARING SSIM FOR IMAGES OBTAINED BY DBIM AND NN

| Case # | DBIM | | NN_map | |
|---|---|---|---|---|
| | $\varepsilon_r$ | σ | $\varepsilon_r$ | σ |
| Fibroglandular | 0.1965 | 0.1129 | 0.7164 | 0.6377 |
| Dense | 0.1652 | 0.0895 | 0.6772 | 0.5981 |
| Very Dense | 0.1255 | 0.0674 | 0.6435 | 0.5593 |

For a quantitative comparison, we used the structural similarity index method (SSIM) to evaluate the quality of the permittivity and conductivity images reconstructed by DBIM and NN. SSIM assesses image similarity based on luminance, contrast, and structure information, yielding a score between 0 and 1, where values closer to 1 indicate higher similarity and better image quality (when compared to a reference). In our evaluation, the ground truth permittivity image served as the reference, and the reconstructed images were assessed accordingly. Table 1 lists the SSIM values for both the DBIM and NN methods, corresponding to the reconstructed images shown in columns (b), (c), (e), and (f) of Fig. 6.

## IV. CONCLUSION

We proposed a fast and interpretable learning-based framework for microwave breast image reconstruction. The system integrates raw signals from conformal antennas and a probability map—generated from antenna spatial coordinates—into an encoder–bottleneck–decoder architecture. Since the conformal antennas directly contact the skin, their positions vary with breast size. To address this, a TM was introduced to model spatial relationships among antennas based on their coordinates. The system demonstrated high efficiency in reconstructing breast tissue profiles, showcasing its potential for real-time medical applications. Compared to DBIM, it provides more detailed tissue patterns with significantly reduced computation time. This approach, which combines the benefits of DL with the flexible, conformal antenna design, opens the door for more accurate and scalable microwave breast imaging systems.